\documentclass[12pt]{article}

\begin{document}

\title{\bf Symmetries of Locally Rotationally Symmetric Models}

\author{M. Sharif \thanks{msharif@math.pu.edu.pk}
\\ Department of Mathematics, University of the Punjab,
\\ Quaid-e-Azam Campus Lahore-54590, PAKISTAN.}

\date{}

\maketitle

\begin{abstract}
Matter collineations of locally rotationally symmetric spacetimes
are considered. These are investigated when the energy-momentum
tensor is degenerate. We know that the degenerate case provides
infinite dimensional matter collineations in most of the cases.
However, an interesting case arises where we obtain proper matter
collineations. We also solve the constraint equations for a
particular case to obtain some cosmological models.
\end{abstract}


\date{}

\newpage

\section{Introduction}

The purpose of this paper is to study matter collineations of
locally rotationally symmetric (LRS) spacetimes. Throughout the
paper $M$ will denote the usual smooth (connected, Hausdorff,
4-dimensional) spacetime manifold with smooth Lorentz metric $g$
of signature $(+,-,-,-)$. Thus $M$ is paracompact. A comma,
semi-colon and the symbol $\pounds$ denote the usual partial,
covariant and Lie derivative, respectively, the covariant
derivative being with respect to the Levi-Civita connection on $M$
derived from $g$. The associated Ricci and stress-energy tensors
will be denoted in component form by $R_{ab}(\equiv R^c{}_{bcd})$
and $T_{ab}$ respectively.

In recent years, much interest has been shown in the study of
matter collineation (MCs) [1]-[9]. A vector field along which the
Lie derivative of the energy-momentum tensor vanishes is called an
MC, i.e.,
\begin{equation} \pounds_{\xi}T_{ab}=0,
\end{equation}
where $\xi^a$ is the symmetry or collineation vector. The MC
equations, in component form, can be written as
\begin{equation}
T_{ab,c} \xi^c + T_{ac} \xi^c_{,b} + T_{cb} \xi^c_{,a} = 0,
\end{equation}
where the indices $a,b,c$ run from $0$ to $3$. Also, assuming the
Einstein field equations, a vector $\xi^a$ generates an MC if
$\pounds_{\xi}G_{ab}=0$. It is obvious that the symmetries of the
metric tensor (isometries) are also symmetries of the Einstein
tensor $G_{ab}$, but this is not necessarily the case for the
symmetries of the Ricci tensor (Ricci collineations) which are
not, in general, symmetries of the Einstein tensor. In a very
recent work, M. Tsamparlis et al. [9] have discussed Ricci and
matter collineations of LRS metrics for the non-degenerate case
only. Here we calculate MCs of hypersurface homogeneous spacetimes
which are locally rotationally symmetric models for degenerate
case and relate them with isometries.

Carot et al. [1] and Hall et al. [2] have noticed some important
general results about the Lie algebra of MCs given in the
following:
\par \noindent
\par \noindent
(i) The set of all MCs on $M$ is a vector space, but it may be
infinite dimensional and may not be a Lie algebra. If $T_{ab}$ is
degenerate, i.e., $det(T_{ab})=0$, then $rank(T_{ab})<4$ and we
cannot guarantee the finite dimensionality of the MCs. If $T_{ab}$
is non-degenerate, i.e., $det(T_{ab})\neq 0$, then
$rank(T_{ab})=4$ and the Lie algebra of MCs is finite dimensional.
\par \noindent
\par \noindent
(ii) If the energy-momentum tensor is of rank $4$ everywhere then
it may be regarded as a metric on $M$. Then it follows by a
standard result that the family of MCs is, in fact, a Lie algebra
of smooth vector fields on $M$ of finite dimension $\leq 10$ (and
$\neq 9$).
\par \noindent
\par \noindent
(iii) Assuming $f$ is a scalar function defined on $M$, then
$\xi=f{\bf X}$ is also an MC if and only if either $f$ is a
constant on $M$ or ${\bf X}$ satisfies $T_{ab}X^a=0$, in which
case $T_{ab}$ is necessarily degenerate and ${\bf X}$ is an
eigenvector of the energy-momentum tensor with eigenvalue $T/2$.
\par \noindent
\par \noindent
(iv) If a vector field $\xi$ on $M$ is a symmetry of {\it all} the
gravitational field sources, then one could require $\pounds_\xi
T_{ab}=0$ (for the {\it non-vacuum} sources) and $\pounds_\xi
C^a_{bcd}=0$ (for the {\it vacuum} sources), where $C^a_{bcd}$ are
Weyl curvature tensor components. This leads to a famous result
given by Hall et al. [2]
\par \noindent
\par \noindent
{\bf Theorem:} Let $M$ be a spacetime manifold. Then, generically,
any vector field $\xi$ on $M$ which simultaneously satisfies
$\pounds_\xi T_{ab}=0$ ($\Leftrightarrow\pounds_\xi G_{ab}=0$) and
$\pounds_\xi C^a_{bcd}=0$ is a homothetic vector field.

If $\xi^a$ is a Killing vector (KV) ( or a homothetic vector),
then $\pounds_{\xi}T_{ab}=0$, thus every isometry is also an MC
but the converse is not true, in general. Notice that
collineations can be proper (non-trivial) or improper (trivial).
Proper MC is defined to be an MC which is not a KV, or a
homothetic vector.

The rest of the paper is organized as follows. In the next
section, we write down MC equations for LRS spacetimes. In section
3, we shall solve these MC equations when the energy-momentum
tensor is degenerate. In section 4, we evaluate MCs for Bianchi
type V metric and finally, a summary of the results obtained will
be presented.

\section{Matter Collineation Equations}

The locally rotationally symmetric spacetimes have many well known
important families of exact solutions of Einstein field equations
and are studied extensively [10]-[13]. They admit a group of
motions $G_4$ acting multiply transitively on 3-dimensional
non-null orbits spacelike ($S_3$) or timelike ($T_3$) and the
isotropy group is a spatial rotation. The metrics of these models
can be written in the forms [10,11]
\begin{equation}
ds^2 = \epsilon[-dt^2+A^2(t)dx^2]-B^2(t)(dy^2+\Sigma^2(y,k)dz^2),
\end{equation}
\begin{equation}
ds^2 = \epsilon[-dt^2+A^2(t)(dx-\Lambda(y,k)dz)^2]-B^2(t)(dy^2
+\Sigma^2(y,k)dz^2),
\end{equation}
\begin{equation}
ds^2 = \epsilon[-dt^2+A^2(t)dx^2]-B^2(t)e^{2x}(dy^2+dz^2),
\end{equation}
with $\epsilon=\pm 1$ and $k=0,\pm 1$, where $\Sigma(y,k)$ is
$\sin y,~y,~\sinh y$ and $\Lambda(y,k)$ is $\cos y,~y^2/2,~\cosh
y$ for $k=+1,0,-1$ respectively. It is mentioned here that the
value of $\epsilon =\pm 1$ differentiates between the static and
non-static cases as can be seen by interchanging the coordinates
$t,x$. We restrict our attention to the non-static case
($\epsilon=-1$) as the results of the static case can be obtained
consequently. The LRS metrics ($\epsilon=-1$) given by Eq.(3) turn
out to be Bianchi types I (BI) or $VII_0$ (B$VII_0$) for $k=0$,
III (BIII) for $k=-1$ and Kantowski-Sachs (KS) for $k=+1$. The LRS
metrics ($\epsilon=-1$) given by Eq.(4) become Bianchi types II
(BII) for $k=0$, VIII (BVIII) or III (BIII) for $k=-1$ and IX
(BIX) for $k=+1$. The LRS spacetime ($\epsilon=-1$) given by
Eq.(5) represents Bianchi type V(BV) or $VII_h$ (B$VII_h$) metric.
A complete classification of Bianchi types I, III and
Kantowski-Sachs spacetimes according to the nature of
energy-momentum tensors [7] has already been given. Here we work
out MCs of BII, BVIII, BIX, and BV spacetimes only for the
degenerate case.

We write down the MC equations for Bianchi types II, VIII and IX
spacetimes given by Eq.(4). The non-vanishing components of Ricci
and energy-momentum tensors are given in Appendix A. Using these,
we can write the MC Eqs.(2) as follows
\begin{equation}
\dot{T}_0\xi^0 + 2T_0\xi^0_{,0} = 0,
\end{equation}
\begin{equation}
\dot{T}_1\xi^0 + 2T_1(\xi^1_{,1}-\Lambda\xi^3_{,1})= 0,
\end{equation}
\begin{equation}
\dot{T}_2\xi^0 + 2T_2\xi^2_{,2} = 0,
\end{equation}
\begin{eqnarray}
(\Lambda^2\dot{T}_1+\Sigma^2\dot{T}_2)\xi^0+2(\Lambda\Lambda'T_1
+\Sigma\Sigma'T_2)\xi^2-2\Lambda T_1(\xi^1_{,3}\nonumber\\
-\Lambda\xi^3_{,3})+2\Sigma^2T_2\xi^3_{,3} = 0,
\end{eqnarray}
\begin{equation}
T_0\xi^0_{,1} + T_1 (\xi^1_{,0}-\Lambda\xi^3_{,0}) = 0,
\end{equation}
\begin{equation}
T_0 \xi^0_{,2} + T_2 \xi^2_{,0} = 0,
\end{equation}
\begin{equation}
T_0\xi^0_{,3}-\Lambda T_1(\xi^1_{,0}-\Lambda\xi^3_{,0})
+\Sigma^2T_2\xi^3_{,0} = 0,
\end{equation}
\begin{equation}
T_1 \xi^1_{,2}-\Lambda T_1 \xi^3_{,2}+T_2\xi^2_{,1} =0,
\end{equation}
\begin{equation}
\Lambda \dot{T}_1 \xi^0 +
\Lambda'T_1\xi^2-T_1(\xi^1_{,3}-\Lambda\xi^3_{,3})+\Lambda
T_1(\xi^1_{,1} -\Lambda\xi^3_{,1})-\Sigma^2 T_2\xi^3_{,1} = 0,
\end{equation}
\begin{equation}
T_2 \xi^2_{,3}-\Lambda
T_1(\xi^1_{,2}-\Lambda\xi^3_{,2})+\Sigma^2T_2\xi^3_{,2} =0,
\end{equation}
where dot  and prime denote differentiation with respect to time
coordinate "t" and $"x"$ respectively. Notice that we have used
the notation $T_{aa}=T_a$. We solve these equations when
$det(T_{ab})=0$. It should be mentioned here that we shall use
$\sqrt{T_0}$ with $T_0>0$ in order to fulfill the dominant energy
condition.

\section{Matter Collineations in the Degenerate Case}

In order to solve MC equations (6)-(15) when at least one
component of $T_m=0,~(m=0,1,2)$, we can have the following two
main cases:
\par \noindent
\par \noindent
(1) when only one of the $T_m \neq 0$,
\par \noindent
\par \noindent
(2) when exactly two of the $T_m \neq 0$,
\par \noindent
\par \noindent
It is mentioned here that the trivial case, where $T_m=0$, shows
that every vector field is an MC.
\par \noindent
\par \noindent
{\bf Case (1)}: This case can further be subdivided into three
cases:
\par \noindent
\par \noindent
(1a) $T_0 \neq 0,\quad T_{j} = 0 \quad (j = 1,2)$,
\par \noindent
\par \noindent
(1b) $T_1 \neq 0, \quad T_{k} = 0 \quad (k = 0,2)$,
\par \noindent
\par \noindent
(1c) $T_2 \neq 0, \quad T_{l} = 0 \quad (l = 0,1)$.
\par \noindent
\par \noindent
The case (1a) is trivial and we get
\begin{equation}
\xi =
\frac{c_0}{\sqrt{T_0}}\partial_t+\xi^i(x^a)\partial_i,~(i=1,2,3),
\end{equation}
where $c_0$ is a constant.

When $T_1=0=T_2$, using values of these components of the
energy-momentum tensor from Eqs.(A2), it follows that
\begin{equation}
-2\frac{A^2\ddot{B}}{B}-\frac{A^2\dot{B}^2}{B^2}
+3\frac{A^4}{4B^4}-k\frac{A^2}{B^2}=0,\quad (k=0,+1,-1),
\end{equation}
\begin{equation}
-B\ddot{B}-\frac{\ddot{A}B^2}{A}-\frac{\dot{A}B\dot{B}}{A}
-\frac{A^2}{4B^2}=0.
\end{equation}
These are second order non-linear differential equations in $A$
and $B$ and can only be solved by assuming some relationship
between these two functions. If we assume that $B=cA$, where $c$
is an arbitrary constant, we obtain
\begin{eqnarray}
2A\ddot{A}+\dot{A}^2+\frac{4kc^2-3}{4c^4}=0,\nonumber\\
2A\ddot{A}+\dot{A}^2+\frac{1}{4c^4}=0.
\end{eqnarray}
The general solution of these equations can not be found
analytically. To have some particular solution, we make an
assumption that $A$ be of the form $A=(at+b)^n$, where $a,~b,~n$
are arbitrary constant. We find that only the possible solution is
for $n=1$. Thus we obtain the following solution
\begin{equation}
ds^2 = dt^2-(at+b)^2(dx-\Lambda(y,k)dz)^2-c^2(at+b)^2(dy^2
+\Sigma^2(y,k)dz^2),
\end{equation}
where $4a^2c^4-4kc^2-5=0$. It can easily be verified that these
metrics represent perfect fluid dust solutions. The energy density
is given by
\begin{eqnarray*}
\rho=\frac{3a^2-4kc^2-1}{4c^4(at+b)^2}.
\end{eqnarray*}

For the case 1(b), it follows from MC Eqs.(6)-(15) that either
$T_1=constant$ or $\xi^0=0$. When $T_1=constant$, we get
\begin{eqnarray}
\xi^0=\xi^0(x^a),~~\xi^1=\Lambda\xi^3+C(y,z),\nonumber\\
\xi^2=\frac{1}{\Lambda'}C_{,3}(y,z),~~\xi^3=-\frac{1}{\Lambda'}
C_{,2}(y,z),
\end{eqnarray}
where $C$ is an integration function of $y$ and $z$. It can be
checked that, in this case, energy density and pressure both
vanish.

In the case of 1(c), solution of the MC equations will become
\begin{equation}
\xi^0=2\frac{T_2}{\dot{T}_2}c_0z\Sigma',~~\xi^1=\xi^1(x^a),
~~\xi^2=-c_0z\Sigma,~~ \xi^3=c_0\int{\frac{dy}{\Sigma}}+c_1,
\end{equation}
where $c_0$ and $c_1$ are arbitrary constants. This case also
gives both pressure and energy density zero for this model. We see
that the case (1) give infinite number of MCs.
\par \noindent
\par \noindent
{\bf Case (2)}: This case implies the following three
possibilities:
\par \noindent
\par \noindent
(2a) $T_l \neq 0, \quad T_2 = 0$,
\par \noindent
\par \noindent
(2b) $T_j \neq 0, \quad T_0 =0$,
\par \noindent
\par \noindent
(2c) $T_k \neq 0, \quad T_1 =0$.
\par \noindent
\par \noindent
The case 2(a) explores further two possibilities i.e. either
$T_1=constant\neq 0$ or $\xi^0=0$. For $T_1=constant$, solution of
the MC equations yields
\begin{equation}
\xi=\frac{c_0}{\sqrt{T_0}}\partial_t+\xi^1(z)\partial_x
+\frac{1}{\Lambda'}\xi^1_{,3}\partial_y.
\end{equation}
When $T_2=0$, using the same procedure as in the case 1(a), we
find the same solution with the condition given as $4a^2c^4+1=0$.

In the case 2(c), from MC Eqs.(6)-(15), in addition to the
improper MCs $\xi_{(1)},\xi_{(2)},\xi_{(3)}$ given in Appendix A,
we obtain the following proper MCs
\begin{eqnarray}
\xi_{(4)}=\xi^1(x^a)\partial_x,\nonumber\\
\xi_{(5)}=\frac{\Sigma \sin z}{\sqrt{T_0}}[\frac{\dot{T_2}}{2k
T_2}{\bf X}_{1}+\partial_t],\nonumber\\
\xi_{(6)}=\frac{\Sigma \cos z}{\sqrt{T_0}}[\frac{\dot{T_2}}{2k
T_2}{\bf X}_{2}+\partial_t],\nonumber\\
\xi_{(7)}=\frac{\Sigma}{\sqrt{T_0}}[-\frac{\dot{T_2}}{2
T_2}\partial_y+\frac{\Sigma'}{\Sigma}\partial_t],
\end{eqnarray}
in which the constraint equation is
\begin{equation}
\frac{T_2}{\sqrt{T_0}}[-\frac{\dot{T_2}}{2
T_2\sqrt{T_0}}\dot{]}=-k.
\end{equation}
The values of ${\bf X}_1$ and ${\bf X}_2$ are given as
\begin{eqnarray}
{\bf X}_1=\frac{\Sigma'}{\Sigma}\partial_y+\frac{\cot
z}{\Sigma^2}\partial_z,~~\nonumber\\
{\bf X}_2=\frac{\Sigma'}{\Sigma}\partial_y+\frac{\tan
z}{\Sigma^2}\partial_z,
\end{eqnarray}
Again we see that all the possibilities of the case (2) give
infinite-dimensional MCs.

\section{Matter Collineations of Bianchi Type V Metric}

The LRS spacetime ($\epsilon=-1$) given by Eq.(5) represents
Bianchi type V metric. MC equations for this metric will become
\begin{equation}
T_{0,0}\xi^0 + 2T_0\xi^0_{,0} = 0,
\end{equation}
\begin{equation}
T_{1,0}\xi^0 + 2T_1\xi^1_{,1}= 0,
\end{equation}
\begin{equation}
\xi^2_{,2}-\xi^3_{,3} = 0,\quad (T_2\neq 0),
\end{equation}
\begin{equation}
T_0\xi^0_{,1}+T_1\xi^1_{,0} = 0,
\end{equation}
\begin{equation}
T_0\xi^0_{,2} + T_2 \xi^2_{,0} = 0,
\end{equation}
\begin{equation}
T_0 \xi^0_{,3} + T_2 \xi^3_{,0} = 0,
\end{equation}
\begin{equation}
T_1\xi^1_{,2}+ T_2\xi^2_{,1} = 0,
\end{equation}
\begin{equation}
T_1 \xi^1_{,3}-T_2 \xi^3_{,1}=0,
\end{equation}
\begin{equation}
\xi^2_{,3}+\xi^3_{,2} =0,\quad (T_2\neq 0).
\end{equation}
Again we shall restrict ourselves only for the degenerate case.
The non-vanishing components of Ricci and energy-momentum tensor
are given in Appendix B. There arises two possibilities for this
case according as (1) when one of the components of
energy-momentum tensor is non-zero and (2) when two of the
components are non-zero. As we have done previously, each of these
two cases can further be divided into three subcases.

The case 1(a) yields the same solution as for the case 1(a) of the
previous section. Using the Einstein field equations for the
perfect fluid matter, we find that the model is pressureless (i.e.
$p=0$) and energy density is given as
\begin{eqnarray*}
\rho=2\frac{\dot{A}\dot{B}}{AB}+\frac{\dot{B}^2}{B^2}
-\frac{1}{A^2}.
\end{eqnarray*}
For the case (1b), it follows from MC equations that
\begin{equation}
\xi=\frac{c_0}{\sqrt{T_1}}\partial_x+\xi^n(x^a)\partial_n,
\quad(n=0,2,3),
\end{equation}
where $c_0$ is a constant.
\par \noindent
\par \noindent
In the case (1c), we obtain the following solution
\begin{equation}
\xi=\xi^l(x^a)\partial_l+c_0\partial_y+c_1\partial_z.
\end{equation}
It is obvious that all the possibilities of the case (1) give
infinite dimensional MCs.

For the case 2(a), when we solve MC equations, we obtain the
following constraint
\begin{equation}
-\frac{T_1}{\sqrt{T_0}}(\frac{\dot{T}_1}{2T_1\sqrt{T_0}}\dot{)}
=\alpha,
\end{equation}
where $\alpha$ is an arbitrary constant which can be positive,
zero or negative.
\par \noindent
\par \noindent
When $\alpha>0$, we have the following solution
\begin{eqnarray}
\xi=c_0\partial_x+c_1\frac{1}{\sqrt{T_0}}(\cos\sqrt{\alpha}x
\partial_t-\frac{\dot{T}_1}{2T_1\sqrt{\alpha}}\sin\sqrt{\alpha}x
\partial_x)\nonumber\\
+c_2\frac{1}{\sqrt{T_0}}(\sin\sqrt{\alpha}x\partial_t
+\frac{\dot{T}_1}{2T_1\sqrt{\alpha}}\cos\sqrt{\alpha}x
\partial_x)\nonumber\\
+\xi^2(x^a)\partial_y+\xi^3(x^a)\partial_z.
\end{eqnarray}
The case $\alpha=0$ yields the solution
\begin{eqnarray}
\xi=c_0\partial_x+c_1\frac{1}{\sqrt{T_0}}(x\partial_t
-(\frac{\dot{T}_1}{4T_1}x^2+\sqrt{T_0}\int{\frac{\sqrt{T_0}}
{T_1}dt})\partial_x)\nonumber\\
+c_2\frac{1}{\sqrt{T_0}}(\partial_t-\frac{\dot{T}_1}
{2T_1}x\partial_x)+\xi^2(x^a)\partial_y+\xi^3(x^a)\partial_z.
\end{eqnarray}
For $\alpha<0$, we obtain
\begin{eqnarray}
\xi=c_0\partial_x+c_1\frac{1}{\sqrt{T_0}}(\cosh\sqrt{\alpha}x
\partial_t-\frac{1}{2\sqrt{\alpha} T_1}(\dot{T}_1+4\alpha T_0)
\sinh\sqrt{\alpha}x\partial_x)\nonumber\\
+c_2\frac{1}{\sqrt{T_0}}(\sinh\sqrt{\alpha}x\partial_t
-\frac{1}{2\sqrt{\alpha} T_1}(\dot{T}_1+4\alpha T_0)
\cosh\sqrt{\alpha}x\partial_x)\nonumber\\
+\xi^2(x^a)\partial_y+\xi^3(x^a)\partial_z.
\end{eqnarray}
Again we see that we obtain infinite dimensional MCs.
\par \noindent
\par \noindent
In the case of 2(b), solving MC equations simultaneously, we
obtain
\begin{equation}
\xi=-2\frac{T_1}{\dot{T_1}}A'(x)\partial_t+A(x)\partial_x
+c_0(\partial_y+\partial_z)+c_1\frac{1}{2}(z\partial_y-y\partial_z).
\end{equation}
\par \noindent
\par \noindent
When we solve MC equations using the constraints of the case 2(c),
it follows that
\begin{equation}
\xi=\frac{c_0}{\sqrt{T_0}}\partial_t+\xi^1(x^a)\partial_x
+c_1(\partial_y+\partial_z)+c_2\frac{1}{2}(z\partial_y-y\partial_z).
\end{equation}
We see that $\xi^1$ is an arbitrary function depending on all four
variables, thus we have infinite dimensional MCs.

\section{Conclusion}

This paper has been devoted to the evaluation of MCs for the LRS
models when the energy-momentum tensor is degenerate. We have
concentrated only for Bianchi types II, VIII, IX and IX spacetimes
given by Eqs.(4) and (5) respectively as the classification of
metrics given by Eq.(3) has already been completed [7]. It is
found that when at least one component of $T_{ab}$ is non-zero
(case (1) of sec. 3), all the possibilities yield infinite
dimensional MCs. If one component of $T_{ab}$ is zero (case (2) of
sec. 3), we again obtain infinite-dimensional MCs. However, in
this case, we have proper MCs, in addition, to the improper MCs.
For this case, the solution of the equation $T_2 = 0$ turns out to
be the perfect fluid dust solution. We also observe that all cases
of the Bianchi type V yield infinite dimensional MCs.

We note that the case in which $T_0\neq0~(T_0>0)$ is the only
surviving component of $T_{ab}$ can always be interpreted as a
dust fluid. In the case when $T_0=0$, we do not have dominant
energy condition instead we have energy density zero.

As we have considered degenerate case, it is natural to expect
infinite dimensional MCs as given by Hall et al. [2]. However, it
may be finite dimensional for some case as given in [7] when we
classify MCs for the LRS models given by Eq.(3). The case 2(b)
(section 3) is left open and it needs to be investigated. It is
expected that this case would provide finite dimensional MCs.
Also, we have obtained constraints on the energy-momentum tensor.
It might be interesting to look for more solutions of the
constraint equations or at least examples should be constructed to
satisfy these constraints.

\renewcommand{\theequation}{A\arabic{equation}}
\setcounter{equation}{0}
\section*{Appendix A}

The surviving components of the Ricci tensor are
\begin{eqnarray}
R_{00} & = & -\frac{1}{AB}(2A\ddot{B}+\ddot{A}B), \nonumber \\
R_{11} & = &A\ddot{A}+2\frac{A\dot{A}\dot{B}}{B}
+\frac{A^4}{2B^4},\nonumber \\
R_{22} & = &B\ddot{B}+\dot{B}^2-\frac{A^2}{2B^2}
+\frac{\dot{A}\dot{B}B}{A}
+k,\nonumber \\
R_{33} & = & \Lambda^2R_{11}+\Sigma^2R_{22},\nonumber\\
R_{13}&=&-\Lambda R_{11}.
\end{eqnarray}
where dot represents derivative w.r.t. time coordinate $t$. The
non-vanishing components of energy-momentum tensor $ T_{ab} $ are
\begin{eqnarray}
T_{00}& = & 2\frac{\dot{A}\dot{B}}{AB}+\frac{\dot{B}^2}{B^2}
-\frac{A^2}{4B^4}+\frac{k}{B^2}, \nonumber \\
T_{11}&=&-2\frac{A^2\ddot{B}}{B}-\frac{A^2\dot{B}^2}{B^2}
+3\frac{A^4}{4B^4}-k\frac{A^2}{B^2},\nonumber \\
T_{22}&=&-B\ddot{B}-\frac{\ddot{A}B^2}{A}-\frac{\dot{A}B\dot{B}}{A}
-\frac{A^2}{4B^2},\nonumber \\
T_{33}&=&\Lambda^2T_{11}+\Sigma^2T_{22},\nonumber\\
T_{13}&=&-\Lambda T_{11}.
\end{eqnarray}
The independent KVs associated with the LRS spacetimes are given
by
\begin{eqnarray}
\xi_{(1)}&=&\sin\phi\partial_\theta +\cot\theta\cos\phi
\partial_\phi,\nonumber\\
\xi_{(2)}&=&\cos\phi\partial_\theta-\cot\theta\sin\phi
\partial_\phi,\nonumber\\
\xi_{(3)}&=&\partial_\phi.
\end{eqnarray}

\renewcommand{\theequation}{B\arabic{equation}}
\setcounter{equation}{0}
\section*{Appendix B}

The surviving components of the Ricci tensor are
\begin{eqnarray}
R_0&=&-\frac{\ddot{A}}{A}-2\frac{\ddot{B}}{B}, \nonumber \\
R_1&=&A\ddot{A}+2\frac{A\dot{A}\dot{B}}{B},\nonumber \\
R_2&=&(B\ddot{B}+\dot{B}^2-\frac{B^2}{A^2}
+\frac{\dot{A}\dot{B}B}{A}=R_3.
\end{eqnarray}
The non-vanishing components of energy-momentum tensor $ T_{ab} $
are
\begin{eqnarray}
T_0&=&2\frac{\dot{A}\dot{B}}{AB}+\frac{\dot{B}^2}{B^2}
-\frac{1}{A^2}, \nonumber \\
T_1&=&-2\frac{A^2\ddot{B}}{B}-\frac{A^2\dot{B}^2}{B^2}+1,\nonumber \\
T_2&=&-(B\ddot{B}+\frac{\ddot{A}B^2}{A}+\frac{\dot{A}
\dot{B}B}{A})e^{2x}=T_3.
\end{eqnarray}

\newpage

\begin{description}
\item  {\bf Acknowledgment}
\end{description}

The author would like to thank the referee for useful and accurate
remarks and suggestions.

\vspace{2cm}

{\bf \large References}

\begin{description}

\item{[1]} Carot, J., da Costa, J. and Vaz, E.G.L.R.: J. Math.
Phys. {\bf 35}(1994)4832.

\item{[2]} Hall, G.S., Roy, I. and Vaz, L.R.: Gen. Rel and Grav.
{\bf 28}(1996)299.

\item{[3]} Carot, J. and da Costa, J.: {\it Procs. of the 6th
Canadian Conf. on General Relativity and Relativistic
Astrophysics}, Fields Inst. Commun. 15, Amer. Math. Soc. WC
Providence, RI(1997)179.

\item{[4]} Yavuz, \.I., and Camc{\i}, U.: Gen. Rel. Grav.{\bf
28}(1996)691;\\
Camc{\i}, U., Yavuz, \.I., Baysal, H., Tarhan,
\.I., and Y{\i}lmaz, \.I.: Int. J. Mod. Phys. {\bf D10}(2001)751.

\item{[5]} Camc{\i}, U. and Barnes, A.: Class. Quant. Grav. {\bf
19}(2002)393.

\item{[6]} Sharif, M.: Nuovo Cimento {\bf B116}(2001)673;
Astrophys. Space Sci. {\bf 278}(2001)447; J. Math. Phys. (2003);\\
Sharif, M. and Sehar Aziz: Gen Rel. and Grav. {\bf 35}(2003).

\item{[7]} Camc{\i}, U. and Sharif, M.: Gen Rel. and Grav. {\bf
35}(2003)97.

\item{[8]} Camc{\i}, U. and Sharif, M.: Class. Quant. Grav. {\bf
20}(2003)2169.

\item{[9]} Tsamparlis, M. and Apostolopoulos, Pantelis S.: Gen.
Rel. and Grav. {\bf 36}(2004)47.

\item{[10]} Ellis, G.F.R.: J. Math. Phys. {\bf 8}(1967)1171.

\item{[11]} Stewart, J.M. and Ellis, G.F.R.: J. Math. Phys. {\bf
9}(1968)1072.

\item{[12]} Ellis, G.F.R. and MacCallum, M.A.H.: Commun. Math.
Phys. {\bf 12}(1969)108;\\ MacCallum, M.A.H.: Commun. Math. Phys.
{\bf 20}(1971)57;\\ Collins, C.B.: Commun. Math. Phys. {\bf
23}(1971)137.

\item{[13]} van Elst, H. and Ellis, G.F.R.: Class. Quantum Grav.
{\bf 13}(1996)1099.

\end{description}

\end{document}